\documentclass[10pt,twocolumn,letterpaper]{article}

\usepackage{cvpr}              

\usepackage[dvipsnames]{xcolor}

\definecolor{cvprblue}{rgb}{0.21,0.49,0.74}
\usepackage[pagebackref,breaklinks,colorlinks,citecolor=cvprblue]{hyperref}
\usepackage{multirow}
\usepackage{xcolor}
\usepackage{colortbl}
\def\confName{CVPR}
\def\confYear{2024}

\definecolor{Gray}{gray}{0.9}
\newcommand{\modelname}{PaperOwl}
\newcommand{\dataname}{M-Paper}
\newcolumntype{a}{>{\columncolor{Gray}}c}
\newcommand*\samethanks[1][\value{footnote}]{\footnotemark[#1]}

\title{mPLUG-PaperOwl: Scientific Diagram Analysis \\ with the Multimodal Large Language Model}

\author{
Anwen Hu\thanks{Equal Contribution.}, Yaya Shi\samethanks[1], Haiyang Xu\thanks{Corresponding Author}, Jiabo Ye, Qinghao Ye, Ming Yan\samethanks[2] \\
 Chenliang Li, Qi Qian, Ji Zhang, Fei Huang \\
Alibaba Group \\
{\tt\small {\{huanwen.haw, shiyaya.syy, shuofeng.xhy, ym119608\}@alibaba-inc.com}}
} 

\begin{document}
\maketitle
\begin{abstract}
Recently, the strong text creation ability of Large Language Models(LLMs) has given rise to many tools for assisting paper reading or even writing. However, the weak diagram analysis abilities of LLMs or Multimodal LLMs greatly limit their application scenarios, especially for scientific academic paper writing. 
In this work, towards a more versatile copilot for academic paper writing, we mainly focus on strengthening the multi-modal diagram analysis ability of Multimodal LLMs. By parsing Latex source files of high-quality papers, we carefully build a multi-modal diagram understanding dataset \dataname. By aligning diagrams in the paper with related paragraphs, we construct professional diagram analysis samples for training and evaluation. \dataname~is the first dataset to support joint comprehension of multiple scientific diagrams, including figures and tables in the format of images or Latex codes. Besides, to better align the copilot with the user's intention, we introduce the `outline' as the control signal, which could be directly given by the user or revised based on auto-generated ones.  Comprehensive experiments with a state-of-the-art Multimodal LLM demonstrate that training on our dataset shows stronger scientific diagram understanding performance, including diagram captioning, diagram analysis, and outline recommendation. The dataset, code, and model are available at \url{https://github.com/X-PLUG/mPLUG-DocOwl/tree/main/PaperOwl}.

\end{abstract}
\section{Introduction}
\label{sec:intro}

The strong text creation ability of the Large Language Model(LLM)\cite{llama, gpt3, vicuna, alpaca} inspires the development of paper-writing copilot recently, such as jenni\footnote{\url{https://jenni.ai/}}. However, existing LLMs or Multimodal LLMs are still not fully competent to assist academic paper writing due to the weak scientific diagram analysis abilities.

\begin{figure}
    \centering
    \includegraphics[width=0.9\linewidth]{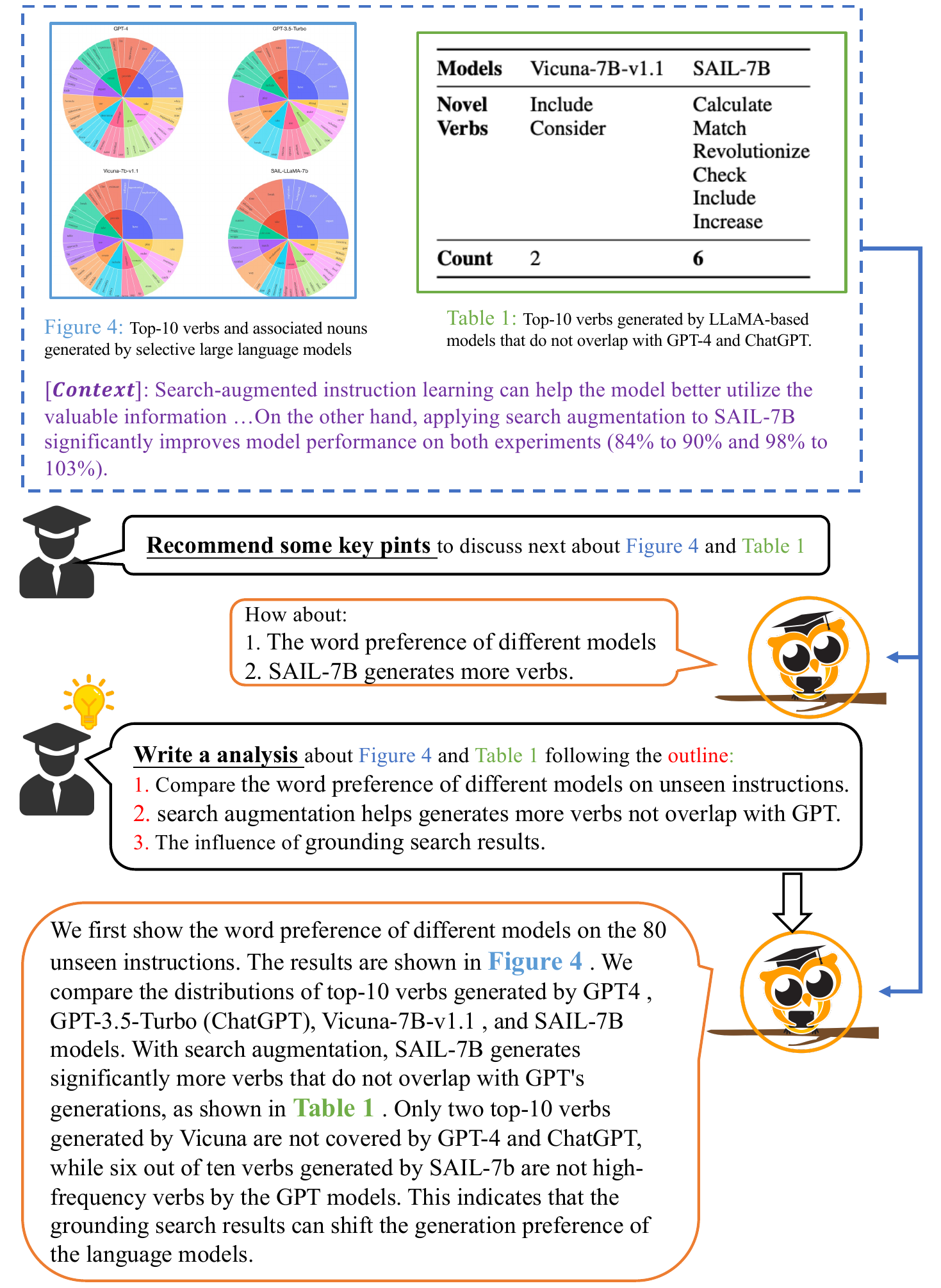}
    \caption{An inllustration of scientific diagram analysis copilot.}
    \label{fig:intro}
\end{figure}

As shown in \cref{fig:intro}, to assist the user in writing academic analysis about scientific diagrams, the copilot should be equipped with major three abilities. \textbf{First and most basically}, the model should be able to understand multiple diagrams of various types (figures, tables, etc.) and in different formats (image or latex). \textbf{Second}, the diagram analysis should remain consistent with the preceding texts and therefore ask to model to correlate multimodal context and diagram information. \textbf{Third}, for better aligning the user's intention, the copilot should be interactable with the user, which requires the model controllable. Recently, there have been many Multimodal Large Language Models(MLLMs)\cite{Alayrac2022FlamingoAV, ye2023mplugowl2, qwenvl, minigpt4, llava, llava1.5, instructblip, cogvlm2023} proposed by connecting a vision encoder with a Large Language Model as the language decoder. These MLLMs are good at chatting about a general image but poor at understanding diagrams. Some work\cite{ureader, docowl} tried to develop MLLMs for Multimodal Document Understanding, covering tables, charts, webpages,.etc. However, these models mainly focus on strengthening the vision comprehension of a single diagram and can't generate detailed scientific analysis.

In this work, to develop scientific diagram analysis skills for the paper-writing copilot, we first build a comprehensive dataset \textbf{\dataname}~to support the learning of the three critical abilities mentioned above. By parsing Latex source files of high-quality academic papers, we carefully extract diagrams in both image and latex formats and align them with their captions and paragraph analysis. To simulate two main scenarios of scientific diagrammatic understanding, we design two main tasks, namely \emph{\textbf{Multimodal Diagram Captioning}} and \emph{\textbf{Multimodal Diagram Analysis}}, where multiple diagrams are the main comprehending objects. In addition, we provide the preceding text, namely $[Context]$, as part of the input to teach the model how to utilize background knowledge and maintain fluency with previous content. Furthermore, to better align users' writing intentions, we design $[Outline]$ as control signals, which are comprised of concise key points to be covered in the analysis. We utilize the ChatGPT to construct $[Outline]$ based on ground-truth paragraph analysis and feed it as the input for \emph{Multimodal Diagram Analysis}. For more user-friendly interaction, recommending $[Outline]$ could inspire users or reduce interaction costs. Thus, we set up another \emph{\textbf{Outline Recommendation}} task to make the copilot more versatile and user-friendly. For accurately evaluating the diagram analysis quality, besides commonly used ngram-based metrics (e.g. CIDEr\cite{cider}), we carefully designed a $\rm{CIDEr}^{gpt}$ score to evaluate both 
n-gram matching and semantic similarity with the help of ChatGPT.

We benchmark multiple state-of-the-art MLLMs on our dataset, validating the challenge of our three tasks. Based on the DocOwl\cite{docowl}, we perform instruction-tuning on a combination of training data from three tasks and propose a strong generalist as the baseline, named \modelname. Comprehensive experiments validate the effectiveness of introducing $[Context]$ and $[Outline]$ as inputs. Besides, we perform sufficient ablation studies about vision encoding to provide insights about the model improvement, such as increasing the image resolution and enhancing the ability to correlate multiple diagrams.

In summary, our contributions are three-fold:
\parskip=0.1em
\begin{itemize}[itemsep=0.1em,parsep=0em,topsep=0em,partopsep=0em]
    \item We build the first high-quality scientific diagram analysis dataset \dataname~to support the learning of correlating multiple diagrams, keeping consistency with the preceding content, and being interactable with users.
    \item Simulating real paper-writing scenarios, we carefully design three multimodal tasks and propose a GPT-based metric, $\rm{CIDEr}^{gpt}$, to measure the paragraph analysis quality by considering both detailed n-gram and overall semantic similarity.
    \item We carefully tune a generalist based on an existing MLLM as the baseline and perform comprehensive experiments to validate the effectiveness of multimodal inputs and training strategies.
\end{itemize}

\section{Related Work}
\label{sec:rela}

\noindent \textbf{Text-only Paper Understanding}\cite{S2ORC2020, AnPaperSum2021, AbuPaperSum2011, SaierF19, Ammar18, ShenMW18} focuses on text and citation graph comprehension in academic papers. 
Such models are competent for a number of text-only thesis comprehension tasks, including information extraction, text classification, paper summarization, or citation recommendation. Benefiting from the strong text understanding ability of Large Language Models(LLMs), many LLM-based tools have been developed as paper-reading assistants, such as ChatDoc\footnote{\url{https://www.chatdoc.com/}}, ChatPDF\footnote{\url{https://www.chatpdf.com/}} and Zhiwen\footnote{\url{https://tongyi.aliyun.com/zhiwen}}.
However, they are still not capable of assisting paper writing due to a lack of multimodal abilities to understand vision information and generate helpful diagram analyses, which are indispensable in scientific papers. 

\noindent \textbf{Multimodal Document Understanding} aims to develop multimodal comprehension abilities for images with rich text information, including charts\cite{chartqa2022, chart2text2022, VisText2023}, tables\cite{wikitableqa, TabFact}, documents\cite{docvqa, mpmqa, deepform, klc} and infographic images\cite{infovqa}, etc. 
In particular, some works\cite{SciGraphQA2023, scicap2021, scicap+2023} focus on understanding scientific figures from papers. Task formats of these work range from Information Extraction\cite{deepform, klc}, Question Answering\cite{docvqa, chartqa2022, infovqa}, Natural Language Inference\cite{TabFact} to Image Captioning\cite{chart2text2022, VisText2023, scicap2021, scicap+2023}. Recently, some works\cite{docowl,ureader, llavar, qwenvl,feng2023unidoc,wang2023tgdoc} have proposed Multimodal Large Language Models with visually-situated text understanding ability. For example, UReader\cite{ureader} performs instruction tuning on an ensembled dataset covering various types of images and designs a Shape-adaptive Cropping Module to process high-resolution document images. However, these MLMMs are still far from acting as a paper-writing copilot for scientific diagram analysis due to main two shortages. First, they can only generate a short answer or description and lack comprehensive diagram analysis abilities. Second, they are all trained to understand a single image, and thus can't correlate context and multiple figures or tables for accurate multimodal analysis. To empower MLMMs with such abilities, we carefully build a scientific diagram analysis dataset \dataname~based on high-quality academic papers. Fineunted on this dataset, our \modelname~shows stronger multimodal diagram analysis abilities and moves a step closer to paper-writing copilot.

\begin{figure*}
    \centering
    \includegraphics[width=1.0\linewidth]{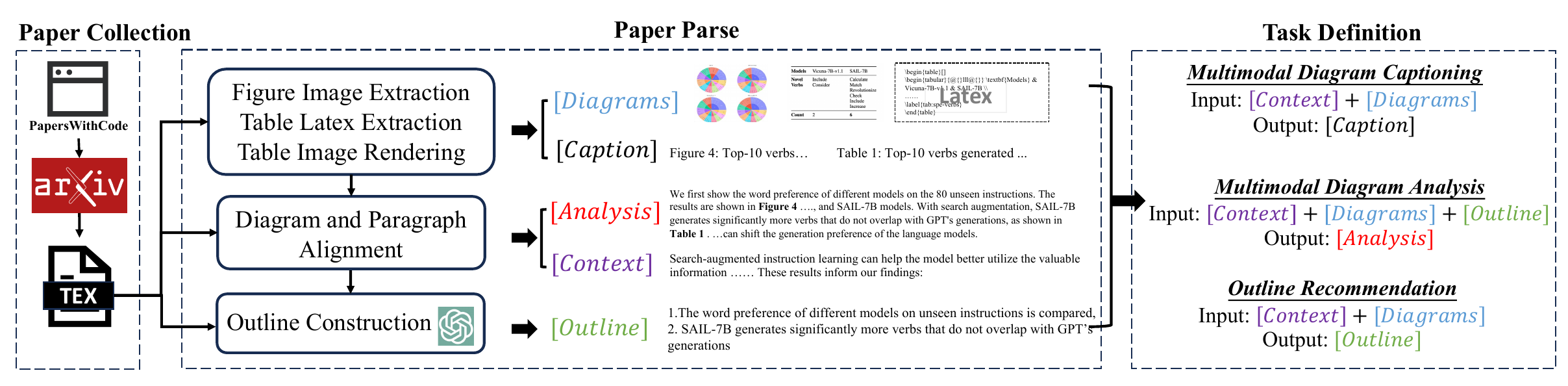}
    \caption{The pipeline  of \dataname~construction and definition of our three tasks.}
    \label{fig:data_process}
\end{figure*}

\section{\dataname}
\label{sec:data}
Towards a paper-writing copilot, this work aims to build \dataname~to help develop multimodal scientific diagram analysis abilities. The dataset construction and task definition are shown in \cref{fig:data_process}.

\subsection{Paper Collection}
The arXiv\footnote{\url{https://arxiv.org/}} is an open-access repository of electronic preprints and postprints, consisting of scientific papers in computer science, mathematics, physics, etc. Due to the field gap, diagrams, writing, and analysis styles are quite different across these fields. In this work, we chose `Computer Science' as the study object. Due to that not all papers are reviewed by peers before posting, the paper quality in arXiv varies a lot and low-quality papers may hurt the model's logical analysis abilities. Considering PapersWithCode\footnote{\url{https://paperswithcode.com/sota}} is a community-driven platform for learning about state-of-the-art research papers on machine learning, we think the quality of papers listed in PapersWithCode is reliable enough. Therefore, with the PapersWithCode API\footnote{\url{https://paperswithcode-client.readthedocs.io/}}, we collect 48k arXiv ids, ranging from 2012 to 2023, covering 15 categories and then download their corresponding Latex source files following official instructions\footnote{\url{https://info.arxiv.org/help/api/basics.html}}.

\subsection{Paper Parse}
PDF and Latex are two kinds of commonly used paper formats in paper-related research. In this work, we choose to parse Latex source files for two main reasons. Firstly, by comparing the content in the `\texttt{$\backslash$ref\{.\}}' tag and `\texttt{$\backslash$label\{.\}}' tag in Latex files, it's easy to accurately align diagrams with paragraph analysis in papers. Secondly, the Latex format is more natural and general for LLM to understand or generate diverse texts, including plain text and mathematical expression, etc. Taking into account these two points, Latex-style text understanding and generation is more suitable for a paper-writing copilot. Following S2ORC\cite{S2ORC2020}, we first parse Latex source files into XML format and then extract diagrams and correlate them with captions and paragraphs. More details on text cleaning can be found in the supplementary material.

\noindent{\textbf{Table Image Rendering.}} Both figures and tables are widely used in scientific academic papers. By parsing the Latext source file, it's easy to align figure reference with figures in image format (e.g.,`jpg') by the `\texttt{$\backslash$includegraphics}' tag. But for tables, there are only Latex codes and no image-format files provided. Towards wider application scenarios, a diagram analysis copilot is necessary to understand tables in both latex and image formats. To support learning such abilities, we further collect table images as inputs. Directly extracting table bounding boxes from PDF-format papers with pdf-parsing tools (e.g., GROBID\footnote{\url{https://github.com/kermitt2/grobid}}) and then cropping table image is a naive way. However, due to the diverse layout in scientific papers, table coordinates given by such tools are not accurate enough. In this work, we collect accurate table images by following three steps. Firstly, we revise the Latex source file to ensure that each table will occupy a separate page after PDF compiling. This operation could greatly reduce the difficulty of table recognition. Then, for each PDF page containing a table, we utilize the classical Edge Detection algorithm Canny\cite{canny1986} to recognize the table bounding box. Finally, the table image is cropped from the PDF page according to the table coordinates. It's worth noting that, to also support the table captioning task and avoid leaking caption information in the cropped table image, the content within the `\texttt{$\backslash$caption\{.\}}' tag is removed during the first step. 

\noindent{\textbf{Outline Construction.}}
During paper writing, for an identical figure or table, even different co-authors can give analysis from different perspectives. Therefore, although a paper-writing copilot can give a comprehensive analysis of a diagram, its analysis can still go against the author's wishes or be inconsistent with the preceding texts. To better cater to users' intentions, we propose to use the `outline' as the intermediate control signal during diagram analysis. Besides directly generating the paragraph analysis, the copilot should also be able to analyze the diagram more accurately following provided key points, namely `outline'. During paper writing, the outline could given by users or generated by the copilot and revised by users. 

For developing such a versatile and controllable copilot, it's necessary to construct appropriate training data for outline generation and analysis generation with outlines. To construct these training samples, in this work, we utilize the GPT-3.5\footnote{\url{https://openai.com/blog/chatgpt}} to generate corresponding outlines for each paragraph by in-context learning. More details can be found in the supplementary material.

\subsection{Task Definition}
After processing Latex source files as mentioned above, we carefully organize these data to support the training and test of multiple tasks designed for the paper-writing copilot, including \emph{Multimodal Diagram Captioning}, \emph{Multimodal Diagram Analysis}, and \emph{Outline Recommendation}.

\noindent{\textbf{Multimodal Diagram Captioning.}}
Different from conventional Image Captioning which aims to describe the attributes and relation between objects, Diagram Captioning requires the model to accurately summarize the content in the figure or table, including some concrete mathematical symbols and proper nouns. Besides, due to partial diagrams being a combination of sub-diagrams, it also asks the model to correlate multiple images. Further, the table during paper-writing can be an image or Latex code, which requires the model to understand different formats of input.

By parsing the Latex source file, it's easy to get diagram captions by extracting content from the `\texttt{$\backslash$caption\{.\}}' tag. For generating captioning more consistent with the paper content and better mentioning prop nouns, we also provide preceding text as the textual input, denoted as $[Context]$. To keep the completeness of semantics, the preceding text is comprised of multiple un-truncated paragraphs before the first reference of the diagram, with max 512 tokens. Thus, the input of Multimodal Diagram Captioning is a triplet of $\langle[Context],  [Diagrams], [Inst]\rangle$, where $[Diagrams]$ can be images of a diagram or Latex code of a table, $[Inst]$ is the instruction. 

Following classical image captioning tasks, we utilize BELU\cite{papineni2002bleu}, METEOR\cite{banerjee2005meteor}, ROUGE-L\cite{lin2004rouge}, and CIDEr\cite{vedantam2015cider} as evaluation metrics. The CIDEr is valued most because it puts higher weight on rarer tokens (e.g., proper nouns), which are more informative.

\noindent{\textbf{Multimodal Diagram Analysis.}}
Much more difficult than writing a caption, Diagram Analysis requires the model to generate a paragraph analysis according to multiple diagrams, even a combination of figures and tables. Besides, diagram analysis is more open-ended than captioning. Different people can analyze a diagram from quite different perspectives. As a paper-writing copilot, the diagram analysis should follow users' intentions as well as possible, otherwise, it will not improve the writing efficiency. Therefore, besides providing the preceding text like the Multimodal Diagram Captioning task to imply the author's intention, we further design the `outline' as the explicit control signal, which instructs key points to discuss with diagrams. Overall, the input of Multimodal Diagram Analysis is a quartet of $\langle[Context],  [Outline], [Diagrams], [Inst]\rangle$.

\begin{figure}
    \centering
    \includegraphics[width=0.9\linewidth]{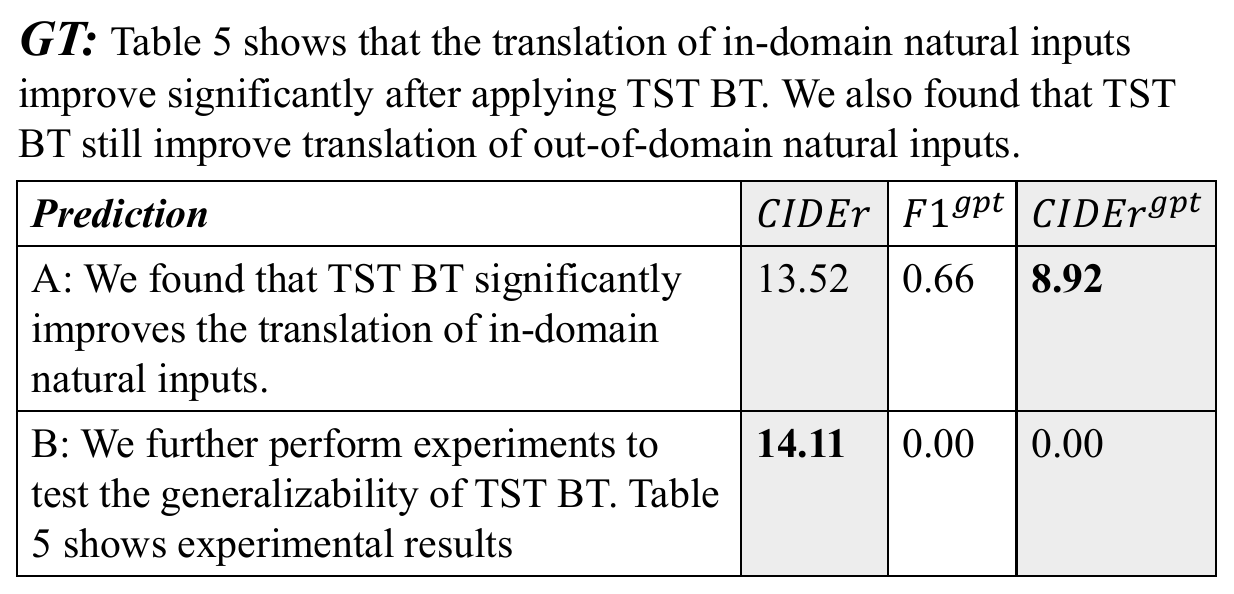}
    \caption{A case of the comparsion of CIDEr and $\rm{CIDEr}^{gpt}$. 
    }
    \label{fig:metric_case}
    \vspace{-10pt}
\end{figure}

Captioning metrics are not quite suitable for paragraph analysis because they mainly measure the n-gram similarity and neglect overall semantic matching. 
To better evaluate the analysis quality, we design a metric to measure the semantic similarity based on GPT 3.5, namely $F1^{gpt}$. Concretely, given the predicted analysis and the ground-truth one, we first prompt the GPT to extract their key points in the list format, respectively. Then, we prompt GPT to judge whether each pair of predicted key point and ground-truth key point matched or not. Finally, we calculate the semantic precision, recall, and F1 score ($F1^{gpt}$) based on GPT's judgment. The detailed prompt design for these two steps can be found in the supplementary material. The $F1^{gpt}$ is good at measuring semantic similarity but hard to assess the quality of detailed descriptions, which is rather what CIDEr is good at. For paragraph analysis, accurately describing key points is more important and we are more tolerant of the form of expression. Considering $F1^{gpt}$ reflects the percentage of mentioning key points and CIDEr measures the n-gram similarity of the whole paragraph. we therefore multiply the CIDEr with $F1^{gpt}$ as the final evaluation metric $\rm{CIDEr}^{gpt}$, where $F1^{gpt}$ plays a critical role. As shown in \cref{fig:metric_case}, prediction A gets a lower CIDEr score because it mentions fewer n-grams within ground truth. However, it describes semantics more accurately and therefore gets a higher $\rm{CIDEr}^{gpt}$ score.

\noindent{\textbf{Outline Recommendation.}}Towards a user-friendly paper-writing copilot, the `outline' can be given directly by users or generated by the copilot and then revised by the user. So recommending outlines accurately is also an important ability for inspiring users or improving writing efficiency. In this work, to develop such ability, we also design an Outline Recommendation task, where the input can be $\langle[Context], [Inst]\rangle$ or $\langle[Context], [Diagrams], [Inst]\rangle$ and the target is $[Outline]$. Captioning metrics are used to evaluate this task.

Instructions of these three tasks can be found in the supplementary material. 

\subsection{Statistic}
\label{sec:statistic}

\noindent\textbf{Paper Category.} 
 \dataname~contains 48,688 papers from more than 15 categories, covering almost all popular research directions in `Deep Learning', especially Computer Vision(CV) and Natural language Processing(NLP). The detailed category distribution can be found in the supplementary material.

\noindent\textbf{Dataset Splits.} \cref{tab:split} shows the split statistic of \emph{Multimodal Diagram Captioning}, \emph{Multimodal Diagram Analysis} and \emph{Outline Recommendation}. For each task, there is no paper overlap across the training, validation and test splits. Both \emph{Multimodal Diagram Captioning} and \emph{Multimodal Diagram Analysis} cover more than 40k papers and provide sufficient training samples. As for \emph{Outline Recommendation}, considering that `outlines' are just intermediate control signals used to interact with users, we don't expect perfect quality of generated outlines. Thus only partial papers are processed to support the training and test of this task.

\begin{table}
\caption{Statistics of \dataname~training, validation and test sets.}
    \label{tab:split}
    \footnotesize
    \tabcolsep=0.2cm
    \centering
    \begin{tabular}{ccccc}
    \toprule
    \textbf{Task} & ~ & \textbf{Train} & \textbf{Val} & \textbf{Test} \\
    \midrule
    Diagram & paper &  46,649 & 479 & 455 \\
    Captioning & sample & 343,546 & 1,131 & 1,133 \\
    \midrule
    Diagram & paper & 40,567 & 412 & 449 \\
    Analysis & sample & 267,476 & 1,087 & 1,195\\
    \midrule
    Outline & paper & 2,548 & 543 & 577 \\
    Recommendation & sample & 78,041 & 3,425 & 3,442 \\
    \bottomrule
    \end{tabular}
\end{table}

\begin{figure}
    \centering
    \includegraphics[width=0.9\linewidth]{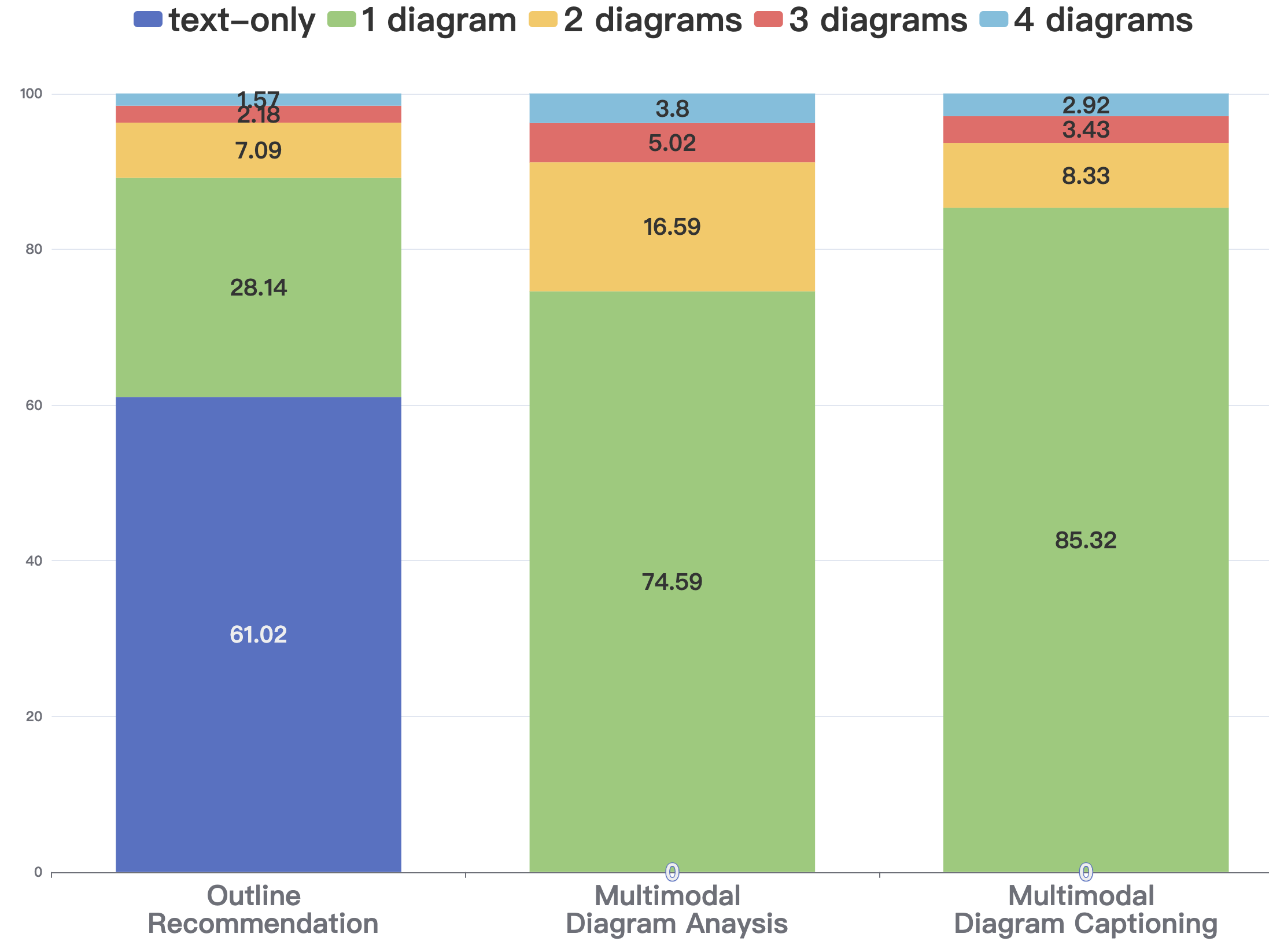}
    \caption{The distribution (\%) of diagram count across 3 tasks. 
    }
    \label{fig:diagram_num_dist}
\end{figure}

\begin{figure}
    \centering
    \includegraphics[width=0.9\linewidth]{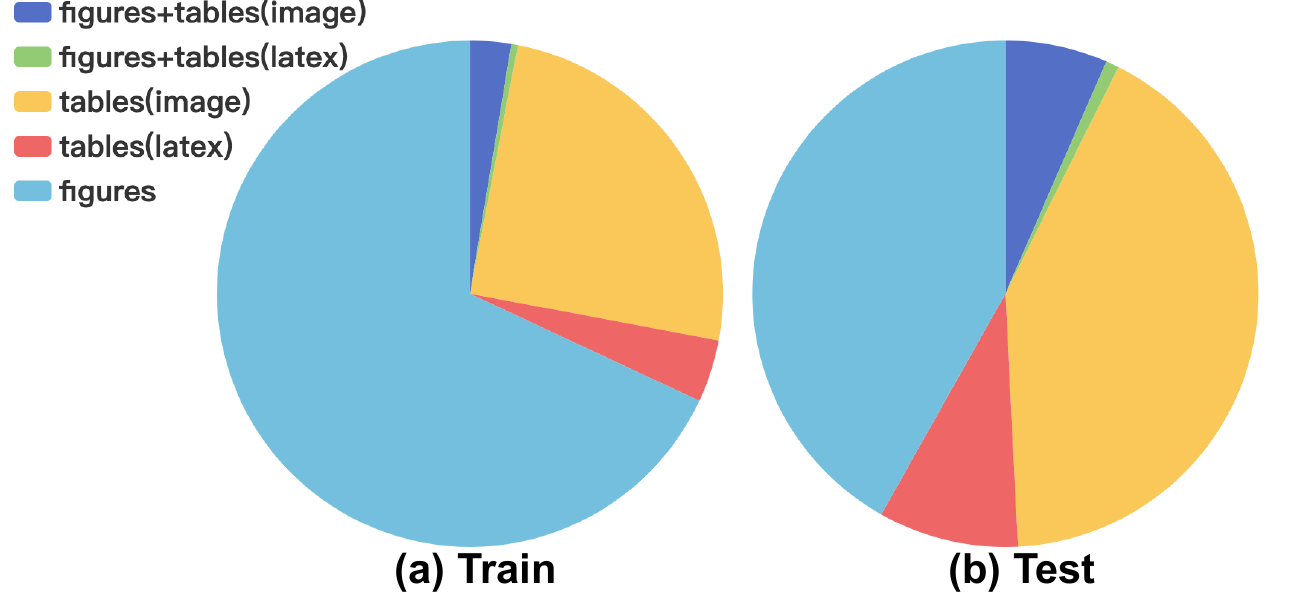}
    \caption{The distribution of diagram types on the training and test set of Multimodal Diagram Analysis. }
    \label{fig:diagram_type_dist}
\end{figure}

\noindent\textbf{Diagram.} As shown in \cref{fig:diagram_num_dist}, the distribution of diagram counts varies across different tasks. For \emph{Multimodal Diagram Analysis}, there are more than 25\% samples with multiple diagrams as inputs, much more than \emph{Multimodal Diagram Captioning}. This indicates that correlating multiple diagrams is a major challenge for \emph{Multimodal Diagram Analysis}. Besides, \cref{fig:diagram_type_dist} shows the distribution of diagram types in \emph{Multimodal Diagram Analysis} task. Our dataset is not limited to a single diagram type but a fusion of figures and tables in the form of vision or latex code. Especially, for better evaluating analysis ability on different diagram types, we slightly balance the diagram type distribution in the test.

\noindent\textbf{Token Length.} \cref{tab:token_len} presents the token length statistic of different textual components in our tasks. The average caption length is much smaller than the paragraph analysis, indicating the \emph{Multimodal Diagram Analysis} task requires both more comprehensive diagram understanding and more detailed description. Besides, the length of the `outline' is far from the `analysis', showing that the input `outline' will not leak too much information about the target analysis but just point out some key points to discuss.  

\begin{table}
\caption{Token length statistic of different textual components.}
    \label{tab:token_len}
    \footnotesize
    \tabcolsep=0.2cm
    \centering
    \begin{tabular}{cccccc}
    \toprule
    ~ & \textbf{Context} & \textbf{Outline} & \textbf{Table Latex} & \textbf{Caption} & \textbf{Analysis} \\
    \midrule
    Mean & 410 & 36 & 177 & 58 & 135  \\ 
    Max & 512 & 126 & 256 & 256 & 256  \\
    \bottomrule
    \end{tabular}
\end{table}

\section{mPLUG-PaperOwl}
\label{sec:model}

\begin{figure}
    \centering
    \includegraphics[width=0.9\linewidth]{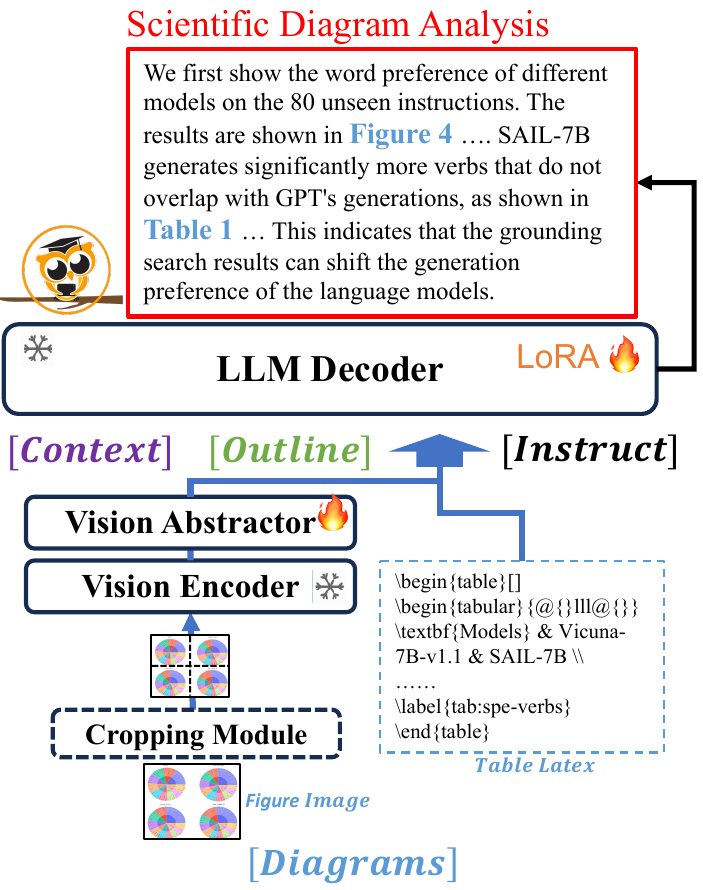}
    \caption{The overall architecture of \modelname.}
    \label{fig:model}
\end{figure}

Existing Multimodal Large Language Models(MLLMs) \cite{mplugowl, minigpt4, llava, qwenvl}  follow a three-module framework, consisting of a vision encoder, a vision-to-text connector, and a Large Language Model as the language decoder. Models with such a framework are easy to adapt to our multimodal tasks by constructing image-text interleaved sequences. 
In this work, we choose one of the state-of-the-art MLLMs: mPLUG-DocOwl\cite{docowl} as the base model to perform instruction-tuning on our \dataname.
 
\subsection{Model Architecture} 
The overall architecture of PaperOwl is shown in \cref{fig:model}.

\noindent{\textbf{Cropping Module.}} Following UReader\cite{ureader}, to better recognize texts in the image, we utilize a parameter-free Cropping Module to cut a 448x448 image to 4 sub-images of 224x224 resolution and then feed each sub-image to the following Vision Encoder independently.

\noindent{\textbf{Vision Encoder.}} The ViT-L/14\cite{vit2021} is utilized as the Vision Encoder, comprised of 24 transformer layers with 16 attention heads and the dimension of hidden states set to 1024. For each image $I$ in the $[Diagrams]$, it's represented as a sequence of visual features $V=\{v_{1}, ...,v_{n}\}$ after the Vision Encoder.

\noindent{\textbf{Vision Abstractor.}} The Vision Abstractor is used to align visual features with the language decoder and aggregate or filter vision semantics. It consists of 6 transformer layers with 8 attention heads and the dimension of hidden states is set as 1024. With 64 learnable tokens $Q=\{q_1,..q_k\}$ as the query, the concatenated sequence $[V:Q]$ as the key and value, the visual features are finally condensed to  $\hat{V}=\{\hat{v}_{1}, ...,\hat{v}_{k}\}$ after cross attention.

\noindent{\textbf{Language Decoder.}} The architecture of Language Decoder is the same as LLaMA-7B\cite{llama}. To adapt to vision-and-language tasks and alleviate catastrophic forgetting, LoRA\cite{hu2022lora} is utilized in the LLM with the rank set as 8. 

\subsection{Model Training}

\noindent{\textbf{Data.}} To develop a versatile paper-writing copilot for scientific diagram understanding, we aim to perform instruction-tuning to enhance an existing MLLM to be a generalist capable of Multimodal Diagram Captioning, Multimodal Diagram Analysis, and Outline Recommendation. Therefore, the training data is a combination of three tasks. Besides, for \emph{Multimodal Diagram Analysis}, to avoid the model heavily relying on `outline' to guess paragraph analysis, samples removing outlines from inputs are also added to the training data to strengthen vision understanding ability. Finally, the total number of instruction-tuning samples is 702,247.

\noindent{\textbf{Details.}} Following most MLLMs\cite{mplugowl, minigpt4, llava}, the Vision Encoder in the PaperOwl is frozen during instruction-tuning to avoid hurting the strong vision representation ability learned during large-scale pretraining. The Vision Abstactro is fine-tuned to better learn how to filter usefully visual diagram information for generating analysis. The raw parameters of LLaMA-7B are frozen, and only the LoRA in the Language Decoder is updated to learn the analysis logic of academic papers. Our model is trained for 10 epochs with the learning rate set as $1e-4$ and the batch size as 256, costing 64 A100 days.

\section{Experiments}
\label{sec:exper}

\begin{table*}
    \caption{The performance comparison with state-of-the-art Multimodal Large Language Models on three tasks. B4, R, M, C and $\rm{C}^{gpt}$ represents BLEU4, ROUGE-L, METEOR, CIDEr and $\rm{CIDEr}^{gpt}$, respectively. `\underline{underline}' means the best zero-shot performance. `Img' refers to the image resolution during training and inference. `Doc' and `Text' refer to using multimodal document and text-only instruction tuning data during training or not.}
    \label{tab:sota_mllm}
    \tabcolsep=0.10cm
    \small
    \centering
    \begin{tabular}{c|ccc|ccca|ccca|ccccca}
    \toprule
    \multirow{2}*{\textbf{Model}} & \multicolumn{3}{c|}{\textbf{Setting}} & \multicolumn{4}{c|}{\textbf{Diagram Captioning}} &  \multicolumn{4}{c|}{\textbf{Outline Recommendation}} &  \multicolumn{6}{c}{\textbf{Diagram Analysis}}  \\
    ~ & Img & Text & Doc &B4 & R & M & C & B4 & R & M & C &B4 & R & M & C & $F1^{gpt}$ & $\rm{C}^{gpt}$ \\ 
    \midrule
    mPLUG-Owl\cite{mplugowl} & 224 & \checkmark & $\times$ & 0.36 & 8.60 & 5.30 & 0.74 & 0.62 & 9.12 & 8.55 & 0.32 &  2.48 & 15.12 & 14.67 & 0.53 & 0.21 & 0.15 \\
    mPLUG-Owl2\cite{ye2023mplugowl2} & 448 & \checkmark & $\times$ &  1.62 & 10.33 & 5.30 & 5.63 & 1.30 & \underline{11.99} & \underline{\textbf{10.48}} & 2.71 & \underline{6.92} & \underline{19.65} & \underline{14.96} & 11.85  & 0.25 & 3.89 \\
    LLaVA 1.5\cite{llava1.5} & 336 & \checkmark & $\times$ & 0.97 & \underline{10.71} & \underline{6.78} & 2.74 & \underline{1.32} & 11.79 & 10.46 & 0.79 & 6.11 & 18.83 & 12.43 & \underline{13.70} & 0.20 & \underline{4.64} \\ 
    Qwen-VL\cite{qwenvl} & 448 & \checkmark & \checkmark & \underline{1.84} & 7.64 & 6.61 & 2.31 & \underline{1.32} & 7.29 & 8.52 & 0.53 & 6.72 & 10.26 & 10.74 & 3.68 & \underline{0.27} & 1.39 \\
    UReader\cite{ureader} & 448 & $\times$ &  \checkmark & 0.56 & 9.84 & 3.34 & 5.95 & 0.25 & 8.17 & 2.88 & \underline{4.59} & 1.22 & 10.59 & 4.33 & 1.02 & 0.05 & 0.05 \\
    DocOwl\cite{docowl} & 448 & \checkmark &  \checkmark & 0.87 & 10.40 & 3.64 & \underline{8.08} & 0.45 & 9.20 & 5.98 & 2.51 & 1.90 & 14.33 & 10.28 & 4.78 & 0.19 & 1.23 \\
    \midrule
    \multicolumn{4}{c|}{PaperOwl} & \textbf{2.37} & \textbf{18.31} & \textbf{7.19} & \textbf{25.50} & \textbf{2.16} & \textbf{17.96} & 7.33 & \textbf{30.65} & \textbf{14.89} & \textbf{30.03} & \textbf{17.56} & \textbf{22.38} & \textbf{0.39} & \textbf{11.50} \\
    \bottomrule
    \end{tabular}
\end{table*}

\subsection{Comparison with SOTA MLLMs.}
We first compare the zero-shot performance of existing MLLMs on our three tasks. As shown in \cref{tab:sota_mllm}, mPLUG-Owl\cite{mplugowl} achieves the worst performance, showing the importance of high resolution for our tasks. After increasing image resolution, mPLUG-Owl2\cite{ye2023mplugowl2} and LLaVA 1.5\cite{llava1.5} outperform the other 3 models trained with multimodal document understanding samples on \emph{Multimodal Diagram Analysis} task. Besides, UReader\cite{ureader}, a model fine-tuned only on document benchmarks, achieves the worst analysis performance. This validates that existing multimodal document understanding data is far from energizing the comprehensive diagram analysis ability of MLLMs and may cause overfitting on question answering or information extraction benchmarks. However, Owl2, LLaVA 1.5 and Qwen-VL all optimize the whole LLM during instruction-tuning while UReader and DocOwl only tune the LoRA. Considering both the performance and training costs, we finally chose DocOwl as our basic model. After fine-tuning with a combination of three tasks, PaperOwl achieves much better performance across three tasks.

\subsection{Ablation Study}
For comprehensively analyzing critical elements for developing a scientific diagram analysis copilot, we perform sufficient comparison experiments to validate the effectiveness of $[Context]$ and $[Outline]$, and present the influence of vision encoding strategies.

\noindent{\textbf{Context Influence.}} For \emph{Multimodal Diagram Captioning} and  \emph{Multimodal Diagram Captioning} tasks, we provide $[Context]$ as auxiliary inputs to implicitly represent users' next writing intention and provide some background information of proper nouns. We first utilize Owl\cite{mplugowl} as the basic model to study whether using $[Context]$ during training and testing. All models are just trained on captioning and analysis tasks and remove $[Outline]$ from inputs. As shown in \cref{tab:context_abla}, for the model trained without $[Context]$, providing $[Context]$ during inference could improve the captioning performance (r2 vs r1), showing $[Context]$ is critical for Diagram Captioning. However, adding $[Context]$ only in testing hurts the analysis performance, indicating the model is hard to balance the comprehension of preceding texts and multiple diagrams for paragraph analysis generation. After adding $[Context]$ in training, the model achieves better performance on both two tasks (r3 vs r2), validating that for better scientific diagram comprehension, it's necessary to incorporate $[Context]$ during both training and inference.

\begin{figure*}
    \centering
    \includegraphics[width=1.0\linewidth]{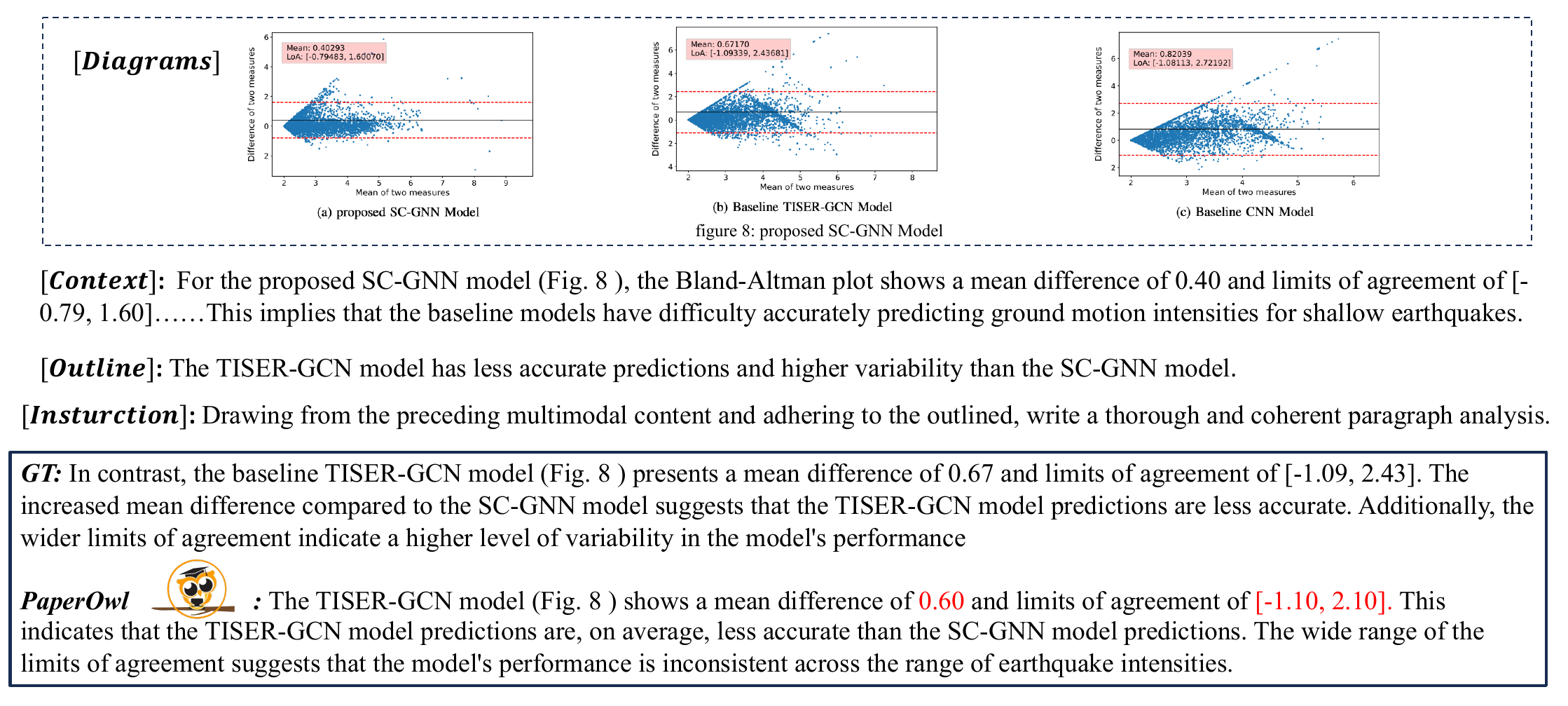}
    \caption{A qualitative result of PaperOwl on Multimodal Diagram Analysis. Wrong descriptions are marked as \color{red}{red}.}
    \label{fig:case}
\end{figure*}

\begin{table}
    \caption{The ablation study about whether utilizing $[Context]$ during training and testing.}
    \label{tab:context_abla}
    \tabcolsep=0.12cm
    \small
    \centering
    \begin{tabular}{c|cc|cca|ccca}
    \toprule
    ~ & \multicolumn{2}{c|}{\textbf{Context}} & \multicolumn{3}{c|}{\textbf{Captioning}} &  \multicolumn{4}{c}{\textbf{Analysis}} \\
    ~ & Train & Test & R & M & C & R & M & C & $\rm{C}^{gpt}$ \\ 
    \midrule
    r1 & $\times$ & $\times$  & 15.43 & 5.45 & 14.67 & 16.56 & 8.71 & 4.45 & 1.47 \\
    r2 & $\times$ & \checkmark  & 16.62 & \textbf{6.82} & 17.72 & 14.44 & 7.66 & 2.87 & 0.94\\
    r3 & \checkmark & \checkmark & \textbf{17.08} & 6.76 & \textbf{21.36} & \textbf{19.25} & \textbf{10.97} & \textbf{7.02} & \textbf{1.81} \\
    \bottomrule
    \end{tabular}
\end{table}

\noindent{\textbf{Outline Influence.}} 
To better align the diagram analysis from a paper-writing copilot with users' intention, we propose to introduce $[Outline]$ as explicit control signals. For validating the effectiveness of $[Outline]$, we further compare variants of Owl about whether utilizing $[Outline]$ during training and testing. As presented in \cref{tab:outline_abla}, for models trained with $[Outline]$ as inputs or not, adding $[Outline]$ during inference could both improve the performance (r2 vs r1, r5 vs r3), showing `Outlines' is an effective control signal for guiding diagram analysis. Besides, even adding pseudo $[Outline]$ generated by the model itself as inputs, the analysis quality could also be improved (r4 vs r3). This indicates that `recommending $[Outline]$ first and then generating diagram analysis' may be a better two-step framework, where the user could also control the copilot by slightly revising the recommended $[Outline]$. Finally, trained with $[Outline]$ makes a significant improvement (r5 vs r2), validating it's essential to teach the model how to correlate multimodal $[Context]$, $[Outline]$ and $[Diagrams]$ for scientific diagram analysis.

\begin{table}
    \caption{The abltion study about the influence of $[Outline]$ for Multimodal Diagram Analysis performance.}
    \label{tab:outline_abla}
    \tabcolsep=0.08cm
    \small
    \centering
    \begin{tabular}{c|cc|ccccaa}
    \toprule
    ~ & \multicolumn{2}{c|}{\textbf{Outline Usage}} & \multirow{2}*{\textbf{B4}} & \multirow{2}*{\textbf{R}} & \multirow{2}*{\textbf{M}} & \multirow{2}*{\textbf{C}} &  &  \\
    ~ & Train & Test & ~ & ~ & ~ & ~ & \multirow{-2}*{\textbf{$F1^{gpt}$}} & \multirow{-2}*{\textbf{$\rm{C}^{gpt}$}} \\ 
    \midrule
    r1 & $\times$ & $\times$ & 6.28 & 19.25 & 10.97 & 7.02 & 0.18 & 1.81 \\ 
    r2 & $\times$ & gpt & 7.23 & 19.86 & 11.24 & 8.99 & 0.22 & 3.10 \\ 
    \midrule
    r3 & gpt & $\times$ & 6.42 & 19.47 & 11.15 & 7.90 & 0.17 & 2.13 \\ 
    r4 & gpt & auto & 5.98 & 19.58 & 11.23 & 9.10 & 0.19 & 2.59 \\ 
    r5 & gpt & gpt & \textbf{15.27} & \textbf{30.36} & \textbf{17.49} & \textbf{21.85} & \textbf{0.41} & \textbf{11.23} \\ 
    \bottomrule
    \end{tabular}
\end{table}

\noindent{\textbf{Vision Encoding Strategies.}}
For vision-and-language tasks, the visual features play a big role in the final performance. In this section, we compare the influence of different vision-representing strategies, including image resolution, whether to fine-tune the Vision Abstractor, and whether to crop the image. As shown in \cref{tab:vision_abla}, during instruction-tuning, freezing the Vision Abstractor greatly hurt the diagram analysis performance (r1 vs r2), validating that fine-tuning the Vision Abstractor is important for adapting an existing MLLM for professional diagram understanding. Besides, at the condition of freezing the Vision Encoder, directly increasing the image resolution and expanding patch position embeddings by bicubic interpolation doesn't bring significant improvement (r3 vs r2), showing that only finetuning the Vsion Abstractor is not enough to adapt to higher-resolution images. 
When equipped with a parameter-free Cropping Module as UReader\cite{ureader} to cut the 448x448 image to 4 sub-images of 224x224 resolutions, the model achieves significantly better performance on the diagram captioning task (r4 vs r2), showing that when the Vision Encoder is frozen, cropping images is a better solution for leveraging higher-resolution images. But, compared with the diagram captioning task, the cropping module still brings a weak improvement to the analysis task. This is mainly because the cropping module results in too many visual tokens (max 1024 tokens from 16 sub-images) and therefore greatly increases the difficulty of multimodal understanding for the language decoder. This shows that how to better encode high-resolution images and balance multimodal inputs is a major challenge for the \emph{Multimodal Diagram Analysis} task.

\begin{table}
    \caption{The ablation study about the training strategy for Multimodal Diagram Captioning and Analysis performance. VA means whether to fine-tune the Vision Abstractor. Crop means whether to use the Croping Module.}
    \label{tab:vision_abla}
    \tabcolsep=0.08cm
    \small
    \centering
    \begin{tabular}{c|ccc|ca|ccca}
    \toprule
    ~ & \multicolumn{3}{c|}{\textbf{Setting}} & \multicolumn{2}{c|}{\textbf{Captioning}} &  \multicolumn{4}{c}{\textbf{Analysis}} \\
    ~ & Img & VA & Crop & M & C  & M & C & $F1^{gpt}$ & $\rm{C}^{gpt}$ \\ 
    \midrule
    r1 & 224 & $\times$ & $\times$  & 5.94 & 23.73  & 16.70 & 18.73 & 0.29 & 8.78 \\
    r2 & 224 & \checkmark & $\times$  & 6.89 & 22.18  & 17.49 & 21.85 & \textbf{0.41} & 11.23 \\
    r3 & 448 &\checkmark & $\times$  & 6.83 & 21.86  & 17.45 & \textbf{22.94} & 0.40 & 11.46 \\
    r4 & 448 & \checkmark & \checkmark  & \textbf{7.19} & \textbf{25.50} & \textbf{17.56} & 22.38 & 0.39 & \textbf{11.50} \\
    \bottomrule
    \end{tabular}
\end{table}

\subsection{Qualitative Results}
\cref{fig:case} presents a qualitative result of \emph{Multimodal Diagram Analysis}. With preceding texts as the input and a simple $[Outline]$ as the control signal, PaperOwl generates a paragraph analysis following the $[Outline]$ and describes more details about diagrams. However, PaperOwl still makes some mistakes about the concrete numbers in the figure, showing the challenge of accurately understanding details among multiple scientific diagrams. More qualitative results of \emph{Multimodal Diagram Captioning} and the comparison of using $[Outline]$ or not can be found in the supplementary material.

\vspace{-5pt}
\section{Conclusion}
To enhance the scientific diagram analysis ability of Multimodal LLMs, we carefully build a multimodal dataset \dataname~based on high-quality Latex files of papers by aligning diagrams with captions and paragraph analysis. Simulating real scenarios of paper writing, we design Multimodal Diagam Captioning, Multimodal Diagram Analysis, and Outline Recommendation tasks. To better evaluate the analysis quality, we propose a GPT-based metric to measure both detailed n-gram matching and overall semantic similarity. We benchmark multiple state-of-the-art MLLMs and propose a strong baseline, PaperOwl, by performing instruction tuning on ensembled training data. Comprehensive experiments validate the effectiveness of the input of the preceding text and outline. Finally, our ablation study provides insights into model improvement, such as increasing image resolution to see more details and how to balance the multimodal information of context, outline and diagrams.
{
    \small
    \bibliographystyle{ieeenat_fullname}
    \bibliography{main}
}

\appendix
\clearpage
\setcounter{page}{1}
\maketitlesupplementary

\section{\dataname}
\subsection{Text Cleaning}
\label{sup:text_clean}
Towards paper-writing copilot, this work focuses on improving the model's multimodal diagram analysis abilities and pays little attention to other writing abilities, such as equation generation or citation recommendation. Both formulas and paper references are virtually impossible to infer from diagrams or preceding texts. Therefore, we further clean paragraph texts by removing such unnecessary information. Concretely, we first replace all citation tags `\texttt{$\backslash$cite\{.\}}' with a special token `\texttt{<cite>}' to remove citation reference. Besides, to avoid generating too long equations, paragraphs containing too long equations ($>40$ chars) in `\texttt{\$.\$}' tags are dropped.

\subsection{Outline Construction}
Taking into account that the `outline' given by users could be multiple content-related key points or a highly concise summary, such as `the overall architecture of our model', we construct two types of outlines by designing different prompts and in-context demonstrations for GPT-3.5, as shown in \cref{tab:simple_summary_prompt} and \cref{tab:detailed_summary_prompt}.

\begin{figure}[!h]
    \centering
    \includegraphics[width=0.8\linewidth]{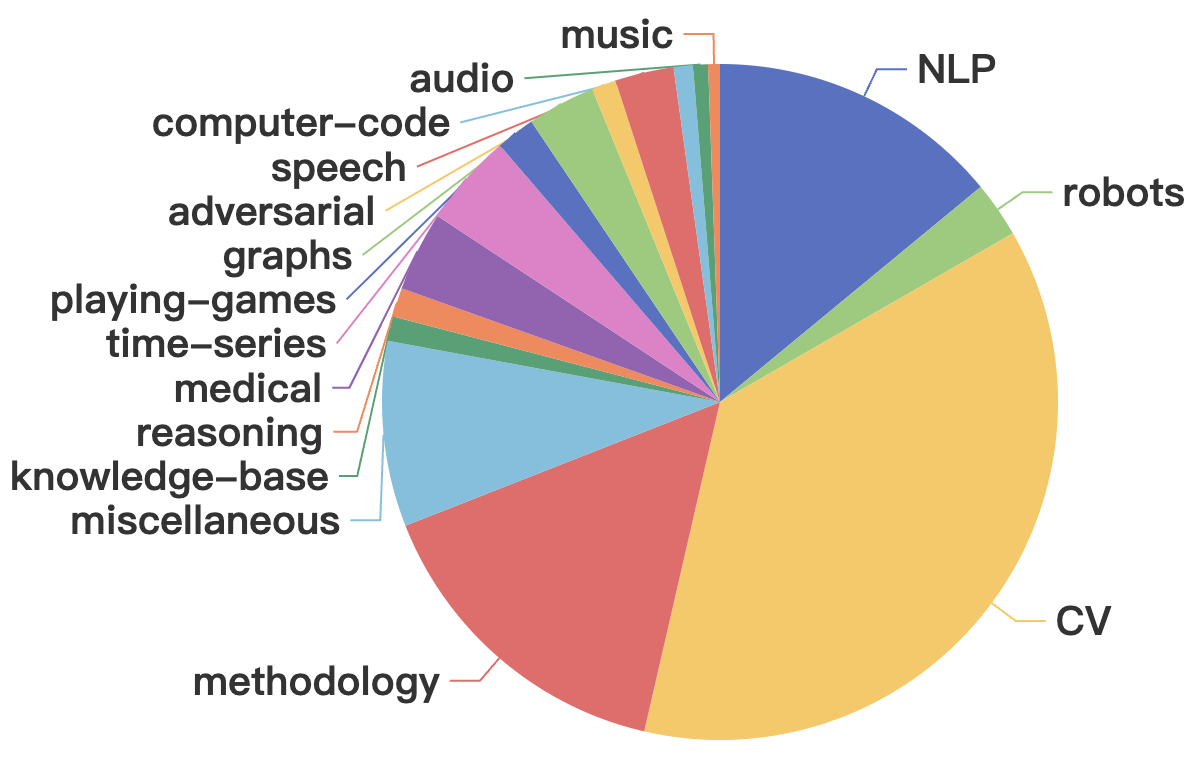}
    \caption{The category distribution of 48,688 academic papers. 
    }
    \label{fig:paper_categ}
\end{figure}

\subsection{Statistic}
The detailed category distribution of papers in \dataname~is shown in \cref{fig:paper_categ}.

\subsection{Task Instruction}
As shown in \cref{tab:instruct_templates}, for each task, we apply various instructions to enhance the model's instruction-following ability.

\section{GPT-based Metric}
For evaluating the overall semantic similarity of a predicted diagram analysis and ground-truth one, we design a GPT-based metric, namely $F1^{gpt}$. We first prompt GPT to extract key points of prediction and ground truth. Then, for each pair of predicted key point and ground-truth one, we further prompt GPT to judge whether it matches or not. Finally, based on GPT's judgments, we calculate the precision, recall, and F1 score ($F1^{gpt}$). The prompts used in these two steps are shown in \cref{tab:gpt_metric}. In particular, during the keypoint extraction process, we prompt GPT to simultaneously process both the prediction and the ground truth to better capture their similarities and differences.

\section{Experiments}

\subsection{Influence of Table Format}
For developing a copilot capable of analyzing different formats of diagrams during paper-writing, \dataname~evaluates table understanding in both image and Latex formats. As shown in \cref{tab:table_abla}, for writing a caption to summarize the table content, understanding Latex is much easier than understanding the image because all data is well-organized in text. However, the Latex format doesn't bring significant improvement for \emph{Multimodal Diagram Anaylysis} and even a decrease in the CIDEr score. This is because when provided latex code of a table, the model tends to describe more rare prop nouns or numbers in the table, which may not be necessary for the discussion and don't appear in the ground-truth analysis. This shows that generating diagram analysis is more challenging at correlating $[Context]$, $[Outline]$, and $[Diagrams]$, rather than mainly understanding the diagram content. 

\begin{table}[!h]
    \caption{The Multimodal Diagram Captioning and Analysis performance on .}
    \label{tab:table_abla}
    \tabcolsep=0.03cm
    \small
    \centering
    \begin{tabular}{c|cca|cccaca}
    \toprule
    \textbf{Table} & \multicolumn{3}{c|}{\textbf{Captioning}} &  \multicolumn{6}{c}{\textbf{Analysis}} \\
    \textbf{Format} & R & M & C  & B4 & R & M & C & $F1^{gpt}$  & $\rm{C}^{gpt}$ \\ 
    \midrule
    Image  & 22.51 & 9.60 & 51.77 & 12.25 & 30.52 & 17.83 & 25.64 & 0.45 & 14.50  \\
    Latex  & 26.69 & 10.54 & 80.03 &  12.03 & 30.38 & 18.11 & 21.56 & 0.47 & 10.51 \\
    \bottomrule
    \end{tabular}
\end{table}

\subsection{More Qualitative Results}
\begin{figure*}
    \centering
    \includegraphics[width=1.0\linewidth]{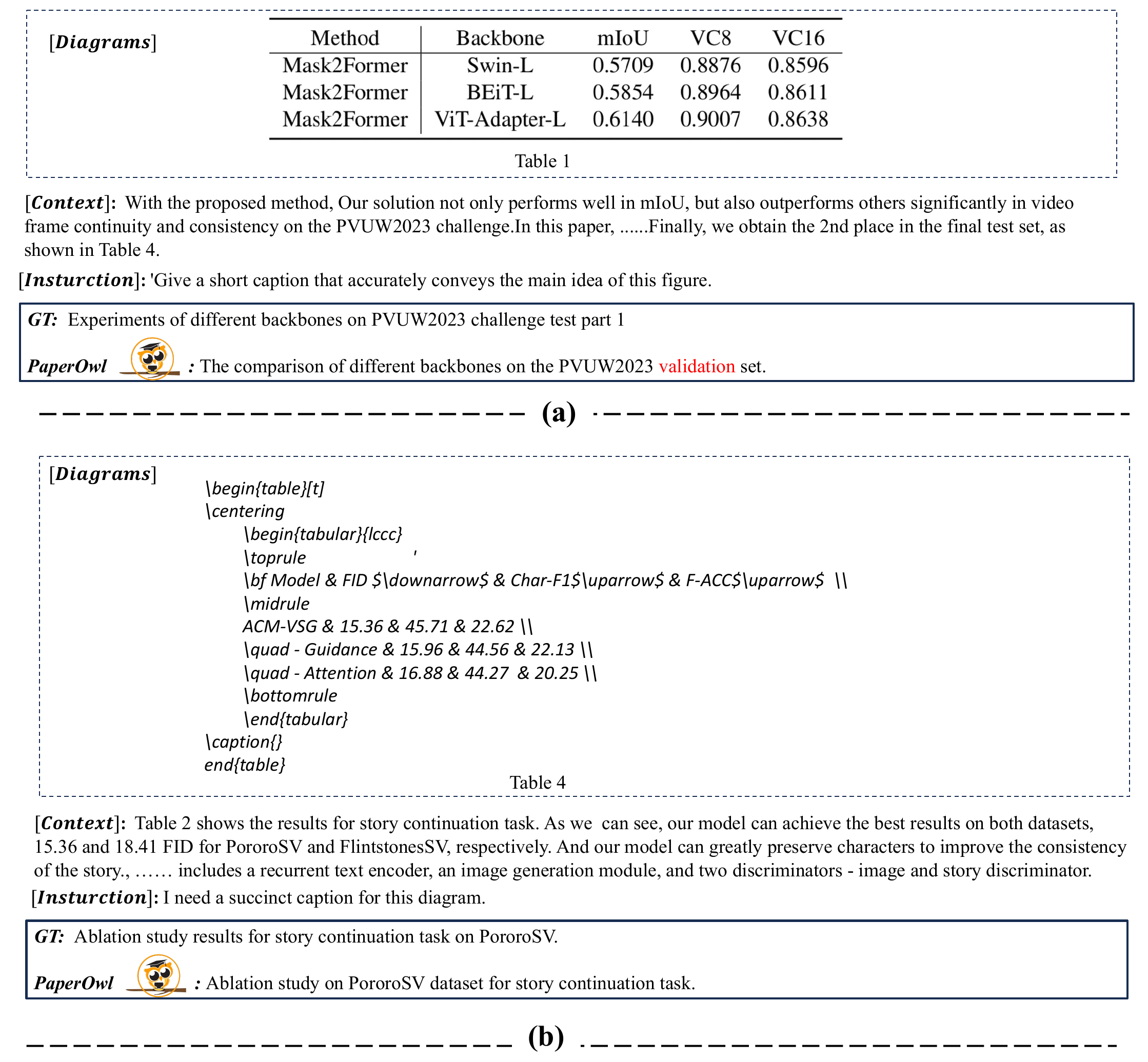}
    \caption{Qualitative results of PaperOwl for captioning tables in the image format (a) and Latex format (b). Wrong descriptions are marked as \color{red}{red}.}
    \label{fig:case_table_cap}
\end{figure*}

\begin{figure*}
    \centering
    \includegraphics[width=1.0\linewidth]{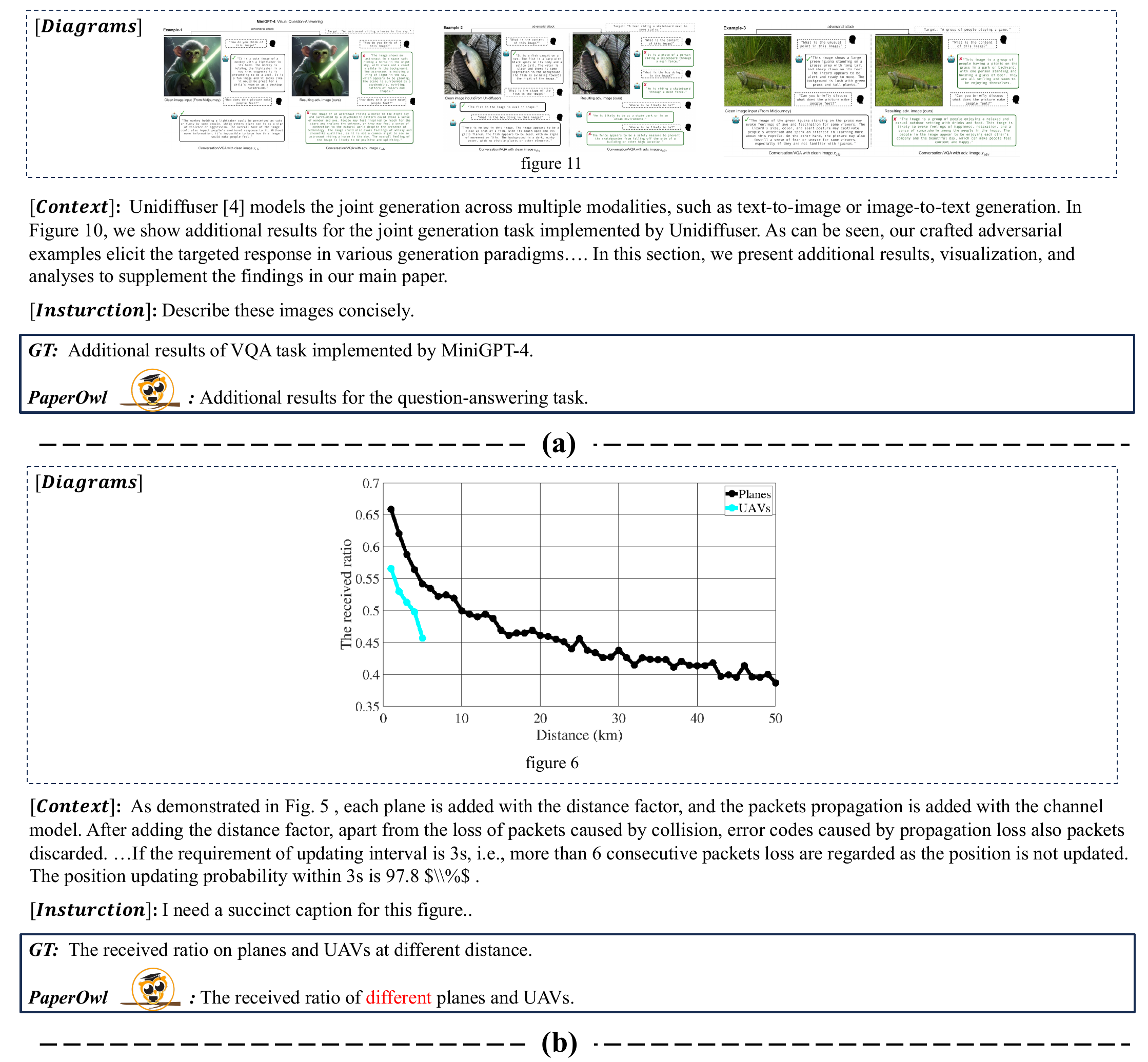}
    \caption{Qualitative results of PaperOwl for captioning figures with multiple sub-images (a) and only 1 image (b). Wrong descriptions are marked as \color{red}{red}.}
    \label{fig:case_figure_cap}
\end{figure*}

\begin{figure*}
    \centering
    \includegraphics[width=1.0\linewidth]{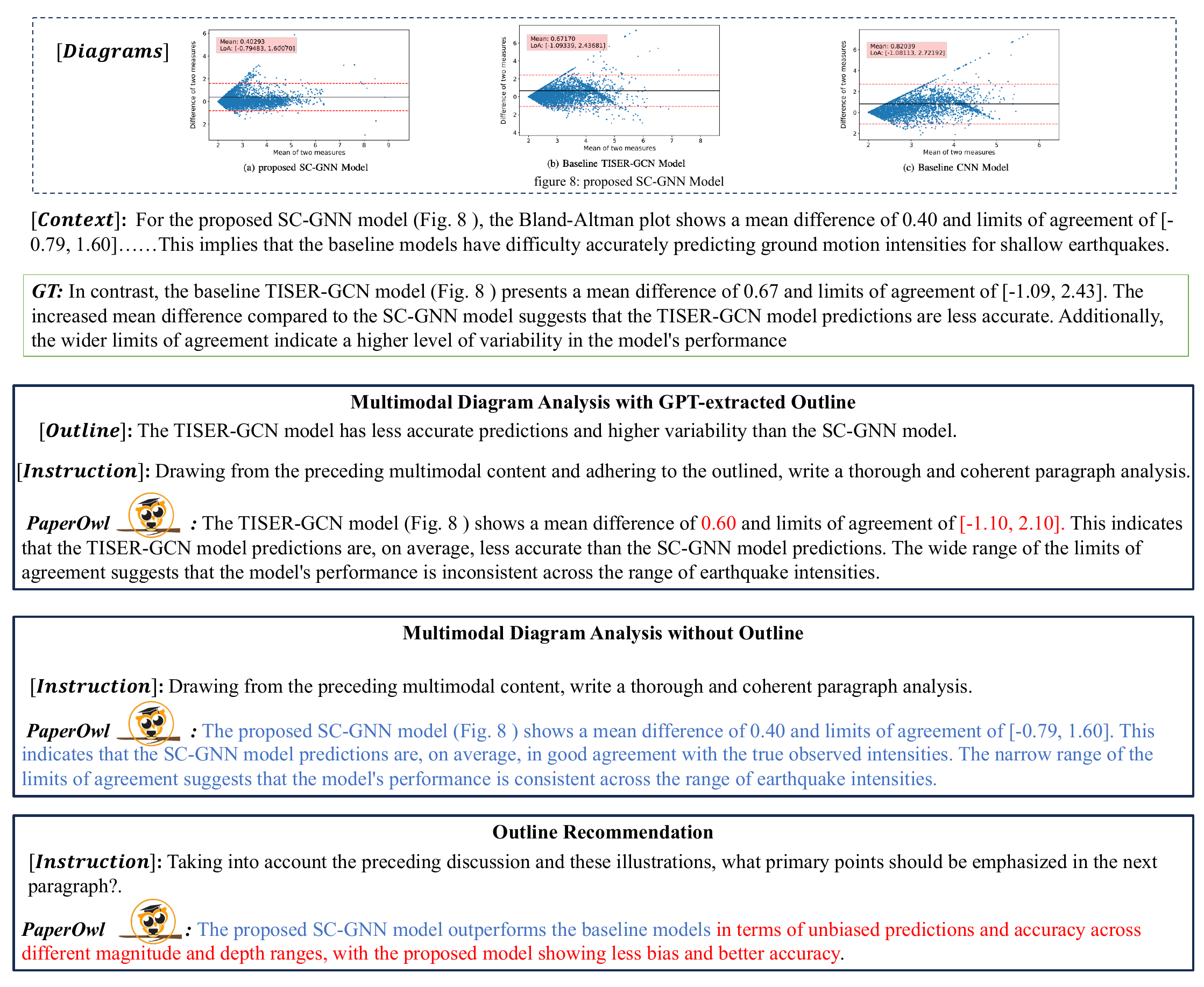}
    \caption{Qualitative results of PaperOwl for Multimodal Diagram Analysis and Outline Recommendation. Wrong descriptions are marked as {\color{red}red}. Descriptions that conform to diagrams but are inconsistent with the ground truth are marked as {\color{blue}blue}.}
    \label{fig:case_figure_analysis}
\end{figure*}

\cref{fig:case_table_cap} and \cref{fig:case_figure_cap} show more qualitative results of Multimodal Diagram Captioning, including multiple sub-figures and tables in the image or latex code format. Besides, as shown in \cref{fig:case_figure_analysis}, without the $[Outline]$, PaperOwl could generate analysis related to diagrams but different from the author's intention. With a concise $[Outline]$, it generates a much better analysis with minor errors, showing the necessity of utilizing $[Outline]$ as the control signal. 

\begin{figure*}
    \centering
    \includegraphics[width=1.0\linewidth]{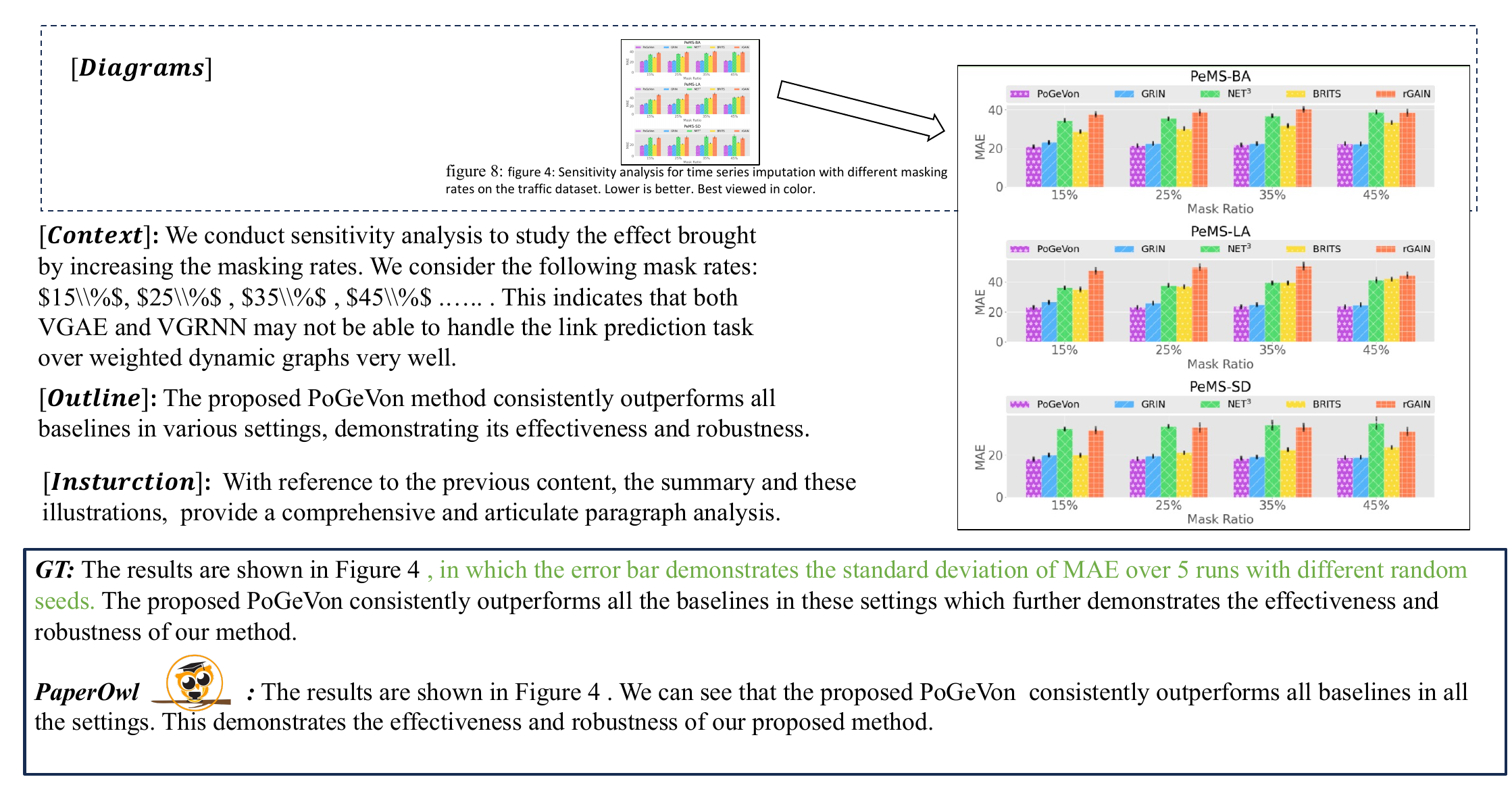}
    \caption{A failure case of PaperOwl for Multimodal Diagram Analysis. Key messages coming from diagrams are marked as {\color{green}green}.}
    \label{fig:case_analysis_bad}
\end{figure*}

As mentioned in \cref{sec:statistic}, during outline construction, the average length of the $[Outline]$ is around 36.43\% of the target diagram analysis to avoid leaking too much information about diagrams. Although we perform such data preprocess to enforce the model learning to describe more diagram details during training, \modelname~sometimes still makes little revisions based on the outline and doesn't provide informative analysis about the diagram, as presented in \cref{fig:case_analysis_bad}. Therefore, it is also a challenge of Multimodal Diagram Analysis to encourage the model to follow the outline while giving more details by understanding diagrams.

\begin{table*}[!hbt]
    \caption{Prompts used for generating a highly concise `outline'.}
    \label{tab:simple_summary_prompt}
    \small
    \centering
    \begin{tabular}{p{16cm}}
    \toprule
    Please provide the main point of the following paragraph which is from a scientific paper. The main point is the central issue in the paragraph and the format like some items in the outline, and it should be as concise and brief as possible!!!! \\
    \\
    Due to the paragraph being from a scientific paper, it can be like: the background of some tasks, or the challenge of previous methods, our methods involve A and B modules, etc for the paragraph from the Introduction section; or experiments results on some datasets for the paragraph from Experiments section, or the pipeline of feature extractor, or the detailed design of some network for the paragraph from Method section. \\
    \\
    Please provide a highly abstract writing purpose for this paragraph like an outline, rather than simply summarizing the content of the paragraph. \\
    \\
    And please generate the main point with less than 20 words! less than 20 words! less than 20 words!!! \\
    \\
    There are some examples of "Paragraph" and "Main Points" pairs. The examples are split by "\#\#\#\#\#\#\#\#\#\#\#\#\#\#\#\#\#\#\#\#\#\#\#\#\#\#\#\#\#\#": \\
    \\
    \#\#\#\#\#\#\#\#\#\#\#\#\#\#\#\#\#\#\#\#\#\#\#\#\#\#\#\#\#\# \\
    Paragraph: \\
    \textbackslash noindent \textbackslash textbf\{Low Reference Dependency\} The Kendall and Spearman correlations between automatic metrics and human judgments with the different numbers of references are shown in Fig.\textbackslash ref\{fig:changing\_reference\_number\}. Our EMScore without any references can achieve competitive results, compared with reference-based metrics which need at least 4 or 5 references, such as BLEU\_1 and Improved\_BERTScore. Besides, our EMScore\_ref with only one reference can achieve comparable results with reference-based metrics, which need at least 8 or 9 references, such as CIDEr and BERTScore. The results show that our metric has lower reference dependency, which benefits from the introduction of video content in evaluation. \\
    \\
    Main Points: \\
    Our metric has a lower reference dependency. \\
    \#\#\#\#\#\#\#\#\#\#\#\#\#\#\#\#\#\#\#\#\#\#\#\#\#\#\#\#\#\# \\
    Paragraph: \\
    Fig.\textbackslash ref\{fig:fine\_grained\_matching\} visualizes how fine-grained EMScore matches the most similar visual elements to the tokens (as the calculation of precision). For the first example, ``bubbles'' occurs in the 106th frame, ``another boy'' occurs in the 160th and 187th frames, and compared with other frames, ``face paint'' appears in a larger proportion in the 4th and 6th frames. For the second example, the visual concept ``boy'' appears as the main visual element in the 53rd frame, so the token 'boy' matches this frame instead of 84th\$\textbackslash sim\$298th frames where multiple visual elements appear. Compared with coarse-grained embedding matching, our fine-grained one can take into account the characteristics of the video, and provide more interpretability for EMScore. \\
    \\
    Main Points: \\
    The visualization results of fine-grained EMScore.\\
    \#\#\#\#\#\#\#\#\#\#\#\#\#\#\#\#\#\#\#\#\#\#\#\#\#\#\#\#\#\# \\
    \\
    Paragraph: $[Paragraph]$\\
    Main Points: $[Main~Points]$\\
    \bottomrule
    \end{tabular}
\end{table*}

\begin{table*}[!hbt]
    \caption{Prompts used for generating an `outline' in the form of multiple key points.}
    \label{tab:detailed_summary_prompt}
    \small
    \centering
    \begin{tabular}{p{18cm}}
    \toprule
    Please use one or several concise sentences to summarize the main points of the following paragraph which is from a scientific paper. \\
    And please note that: \\
    (1) Each sentence should strive to express one main point as succinctly as possible. \\
    (2) Please summarize the most critical points, preferably no more than 3. And one main point is enough for some short paragraphs!!! \\
    (3) If there are multiple main points, use ``1. 2. 3." to list them and use ``\textbackslash n" to split them. \\
    \\
    There are some wrong formats with prefix like this:
    ``The article introduces xxx". \\
    ``The authors conduct experiments xxx". \\
    ``They introduce xx". \\
    ``xxx proposed by the author". \\
    Please directly generate the key points of the paragraph, and don't use the prefix like above. \\
    \\
    There are some examples of "Paragraph" and "Main Points" pairs. The examples are split by "\#\#\#\#\#\#\#\#\#\#\#\#\#\#\#\#\#\#\#\#\#\#\#\#\#\#\#\#\#\#": \\
    \\

    \#\#\#\#\#\#\#\#\#\#\#\#\#\#\#\#\#\#\#\#\#\#\#\#\#\#\#\#\#\# \\
    Paragraph: \\
    Video Captioning\textbackslash cite\{DBLP:journals/tcsv/DengLZWZH22\} aims to generate a text describing the visual content of a given video. Driven by the neural encoder-decoder paradigm, research in video captioning has made significant progress~\textbackslash cite\{DBLP:conf/iccv/VenugopalanRDMD15, DBLP:conf/cvpr/ZhangSY0WHZ20\}. To make further advances in video captioning, it is essential to accurately evaluate generated captions. The most ideal metric is human evaluation while carrying human judgments is time-consuming and labor-intensive. Thus, various automatic metrics are applied for video caption evaluation. \\
    \\
    Main Points: \\
    Accurately evaluating the generated descriptions is necessary, and due to the time-consuming and labor-intensive nature of human judgments, automatic evaluation metrics are widely used. \\
    \\
    \#\#\#\#\#\#\#\#\#\#\#\#\#\#\#\#\#\#\#\#\#\#\#\#\#\#\#\#\#\# \\

    Paragraph: \\
    However, most of the widely applied video caption metrics like BLEU\textbackslash cite\{DBLP:conf/acl/PapineniRWZ02\}, ROUGE\textbackslash cite\{lin-2004-rouge\}, CIDEr\textbackslash cite\{7299087\}, and BERTScore\textbackslash cite\{DBLP:conf/iclr/ZhangKWWA20\} come from the other tasks, such as machine translation, text summarization and image captioning, which may neglect the special characteristic of video captioning and then limit the development of video captioning. Furthermore, these automatic metrics require human-labeled references --- and thus they are called reference-based metrics --- and such requirements cause three intrinsic drawbacks: (1) They can not be used when provided videos have no human-labeled references, which is not uncommon in this age that millions of reference-free videos are produced online every day. (2) They may over-penalize the correct captions since references hardly describe all details of videos due to the one-to-many nature\textbackslash cite\{DBLP:conf/acl/YiDH20\} of captioning task, especially when the number of references is limited. Fig.\textbackslash ref\{fig:introductionexample\} (a) shows one such example where a candidate caption correctly describes the ``a rock'' while reference-based metrics punish this word since references do not contain it. (3) As pointed by \textbackslash cite\{rohrbach-etal-2018-object\}, these reference-based metrics may under-penalize the captions with ``hallucinating'' descriptions since these metrics only measure similarity to references, and the visual relevance cannot be fully captured. For example, as shown in Fig.\textbackslash ref\{fig:introductionexample\} (b), due to the word ``games'' appearing in the references, some reference-metrics return higher scores for caption B than caption A, even though ``different games'' is a ``hallucinating'' phrase which is not related to the video. \\
    \\
    Main Points: \\
    1. Commonly used video caption metrics come from other tasks and may not fully capture the unique characteristics of video captioning. \\
    2. The requirement of reference causes three intrinsic drawbacks: (1) Cannot be applied in real time. (2) Over-penalize the correct captions. (3) Under-penalize the captions with ``hallucinating'' descriptions.

    \#\#\#\#\#\#\#\#\#\#\#\#\#\#\#\#\#\#\#\#\#\#\#\#\#\#\#\#\#\# \\
    \\
    Paragraph: $[Paragraph]$\\
    Main Points: $[Main~Points]$\\

    \bottomrule
    \end{tabular}
\end{table*}


\begin{table*}[!hbt]
    \caption{Instructuion used for Multimodal Diagram Captioning, Multimodal Diagram Analysis and Outline Recommendation. The $[object]$ is randomly chosen from $\{figures, images, photos, pictures, diagrams, illustrations\}$ or $\{figure, image, photo, picture, diagram, illustration\}$ depending on the number of diagrams is more than 1 or not.}
    \label{tab:instruct_templates}
    \small
    \centering
    \begin{tabular}{p{18cm}}
    \toprule
    \multicolumn{1}{c}{\textbf{Multimodal Diagram Captioning}}  \\
    \midrule
    Describe $[object]$ concisely. \\
    Write a caption of $[object]$. \\ 
    Provide a brief description of $[object]$. \\
    Write a short caption for $[object]$. \\
    come up with a concise caption that captures the essence of $[object]$. \\
    Encapsulate the key information presented in $[object]$ in a brief statement. \\
    I need a succinct caption for $[object]$. \\
    Please provide a pithy summary of $[object]$ that effectively communicates its message. \\ 
    Can you provide a snappy caption that perfectly encapsulates the message conveyed by $[object]$? \\
    Please write a brief but compelling caption that grabs the reader's attention and draws them into $[object]$. \\
    Give a short caption that accurately conveys the main idea of $[object]$. \\
    \midrule
    \multicolumn{1}{c}{\textbf{Multimodal Diagram Anaysis}} \\ 
    \midrule
     Based on the previous content and the outline, write a detailed and fluent paragraph analysis. \\
    With reference to the preceding content and the given summary, compose a comprehensive and articulate paragraph analysis. \\
    Considering the information provided earlier and following the provided outline, produce a detailed and fluent analysis in paragraph form. \\
    Drawing from the preceding content and adhering to the outlined structure, write a thorough and coherent paragraph analysis. \\ 
    Based on the previous content and guided by the summary, construct a detailed and fluid analysis in paragraph format. \\
    Taking into account the preceding information and following the provided outline, generate a comprehensive and well-developed paragraph analysis. \\
    Considering the content discussed earlier and following the provided outline, present a detailed and fluent analysis in paragraph form. \\
    With reference to the previous content and the summary, provide a comprehensive and articulate paragraph analysis. \\
    Based on the preceding discussion and in accordance with the outlined structure, compose a detailed and coherent paragraph analysis. \\
    Considering the information presented earlier and adhering to the provided summary, formulate a thorough and seamless paragraph analysis. \\
    \midrule
     \multicolumn{1}{c}{\textbf{Outline Recommendation}} \\
    \midrule
    \emph{more than 1 diagrams} \\
    \midrule
    Based on the previous content and $[object]$, list some key points that should be covered in the next paragraph. \\
    Considering the preceding text with $[object]$, the next paragraph needs to address these essential aspects. \\
    Drawing from the preceding text and image information, what crucial points should be focused on in the ensuing paragraph? \\
    Given the multimodal information provided earlier, write some key factors for the next paragraph. \\
    With reference to the previous discussion and $[object]$, the next paragraph should discuss the following important elements. \\
    In light of the preceding content with $[object]$, which significant points should be analyzed in the subsequent paragraph? \\
    Based on the previous text and $[object]$, the next paragraph should delve into these core aspects. \\
    Considering the text and vision information presented before, give some main factors that should be addressed in the ensuing paragraph. \\
    Taking into account the preceding discussion and $[object]$, what primary points should be emphasized in the next paragraph? \\
    Given the previous context with $[object]$, generate some key elements that should be discussed in the next paragraph should discuss. \\
    \midrule
    \emph{no diagrams} \\
    \midrule
    Based on the previous content, list some key points that should be covered in the next paragraph. \\
    Considering the preceding text, the next paragraph needs to address these essential aspects. \\
     Drawing from the preceding information, what crucial points should be focused on in the ensuing paragraph? \\
     Given the information provided earlier, write some key factors for the next paragraph. \\
     With reference to the previous discussion, the next paragraph should discuss the following important elements. \\
     In light of the preceding content, which significant points should be analyzed in the subsequent paragraph? \\
     Based on the previous text, the next paragraph should delve into these core aspects. \\
     Considering the information presented before, give some main factors that should be addressed in the ensuing paragraph. \\
     Taking into account the preceding discussion, what primary points should be emphasized in the next paragraph? \\
     Given the previous context, generate some key elements that should be discussed in the next paragraph should discuss. \\

    \bottomrule
    \end{tabular}
\end{table*}

\begin{table*}[!hbt]
    \caption{Prompts used for calculate $F1^{gpt}$. $[Prediction]$ and $[Ground~Truth]$ are predicted analysis and ground-truth analysis, respectively. $[Predicted~Point]$ and $[GT~Point]$ is a pair of key points extracted from the $[Prediction]$ and $[Ground~Truth]$, respectively,}
    \label{tab:gpt_metric}
    \small
    \centering
    \begin{tabular}{p{16cm}}
    \toprule
    \multicolumn{1}{c}{\textbf{Prompt GPT for Extracting Key Points}}  \\
    \midrule
    Please summarize the main points of the prediction and ground truth. 
    And strictly with the format: \\ 
    1. xxx.\\
    2. xxx.\\
    ... \\ 
    Please ensure that the generated main points comprehensively condense the information of the original text (prediction or ground truth).
    The number of generated main points can be as many as possible, but no more than 10. \\
    \\ 
    If there are parts of the prediction and ground truth that are the same, reflect that in main points, such as some main points of them are the same, and other main points summarize the unique content of themselves. \\
    \\ 
    Please note that if there are any overlapping contents between the prediction and ground truth, the main points for these contents should remain consistent. However, for different content of them, please provide separate main points for each. \\
    \\
    The format is as follows: \\
    $\#\#\#\#\#\#\#$
    \\
    Predicted text: xxx. \\
    \\
    Ground Truth text: xxx. \\
    \\
    The main points of the predicted text: \\
    1. xx \\
    2. xx \\
    ... \\
    \\
    The main points of the ground truth text: \\
    1. xx \\
    2. xx \\
    ... \\
    $\#\#\#\#\#\#\#$ \\
    \\
    Now, please generate the main points of the given prediction and ground truth, please strictly use the prompt 'The main points of the xxx' in the response. \\
    \\
    Predicted text: $[Prediction]$ \\ 
    Ground Truth text: $[Ground~Truth]$ \\
    \midrule
    \multicolumn{1}{c}{\textbf{Prompt GPT for Judging Semantic Matching}}  \\
    \midrule
    Given a predicted text and a reference text, please judge whether the semantics of the predicted text can match the reference text. \\
    And use Yes or No to represent match or mismatch. \\
    The format is as follows: \\
    Predicted text: xxx. \\
    Reference text: xxx. \\
    Yes/No \\
    ---------- \\
    Predicted text: $[Predicted~Point]$ \\
    Reference text: $[GT~Point]$ \\
    \bottomrule
    \end{tabular}
\end{table*}

\end{document}